\documentclass[%
reprint,
superscriptaddress,
longbibliography,
preprintnumbers,
nofootinbib,
amsmath,amssymb,
floatfix,
aip,
author-year,
]{revtex4-2}

\usepackage[english]{babel}

\usepackage[utf8]{inputenc}
\usepackage{amsthm,amsmath,amssymb}
\usepackage{graphicx}

\usepackage[table]{xcolor}

\usepackage{tikz}
\usetikzlibrary{quotes}
\usetikzlibrary{shapes.misc}

\definecolor{Dark2-A}{RGB}{ 27, 158, 119}
\definecolor{Dark2-B}{RGB}{217,  95,   2}

\definecolor{Set1-A}{RGB}{228,  26,  28}
\definecolor{Set1-B}{RGB}{ 55, 126, 184}

\tikzstyle{e}=[thick,arrows={-latex},rounded corners]
\tikzset{every label/.style={rectangle,fill=none,draw=none, label distance=0pt}}

\tikzstyle{n}=[circle,draw=white,line width=2pt,outer sep=0pt,inner sep=0pt, minimum size=4ex]
\tikzstyle{inline}=[anchor=base]
\tikzstyle{open}=[fill=Dark2-A!50]
\tikzstyle{closed}=[fill=Dark2-B!50]
\tikzstyle{conditioned}=[draw=black,line width=2pt]
\tikzstyle{unfair}=[text=Dark2-B]
\tikzstyle{bias}=[draw=Dark2-B]
\tikzstyle{unobserved}=[fill opacity=0.5]

\newcommand\innode[ 2 ]{ 
  \begin{tikzpicture}[baseline] 
      \node[inline,inner ysep=0pt,inner xsep=3pt,outer sep=0pt,opacity=0] (n) {#1}; 
      \useasboundingbox (n.north west) rectangle (n.south east);
      \node[n,minimum size=3ex,inner sep=1pt,inline,#2] {#1}; 
  \end{tikzpicture}
  }

\makeatletter
\DeclareRobustCommand{\rvdots}{%
  \vbox{
    \baselineskip4\p@\lineskiplimit\z@
    \kern-\p@
    \hbox{.}\hbox{.}\hbox{.}
  }}
\makeatother

\usepackage[colorlinks=true,%
 linkcolor = {Set1-B},%
 citecolor = {Set1-B},%
 urlcolor = {Set1-B},
 unicode]{hyperref}
\usepackage{xurl}
\usepackage{orcidlink}

\usepackage{todonotes}

\graphicspath{ {figures/} }

\makeatletter
\renewcommand*{\@textcolor}[3]{%
  \protect\leavevmode
  \begingroup
    \color#1{#2}#3%
  \endgroup
}
\makeatother
\newcommand\rightbias{\textcolor{Dark2-B}\rightarrow}

\begin{document}

\title{Causal foundations of bias, disparity and fairness}

\author{V.A. Traag\,\orcidlink{0000-0003-3170-3879}}
\email{v.a.traag@cwts.leidenuniv.nl}
\affiliation{Centre for Science and Technology Studies (CWTS), Leiden University, the Netherlands}

\author{L. Waltman\,\orcidlink{0000-0001-8249-1752}}
\affiliation{Centre for Science and Technology Studies (CWTS), Leiden University, the Netherlands}

\date{\today}

\begin{abstract}
  The study of biases, such as gender or racial biases, is an important topic in the social and behavioural sciences.
  However, the literature does not always clearly define the concept.
  Definitions of bias are often ambiguous or not provided at all.
  To study biases in a precise manner, it is important to have a well-defined concept of bias.
  We propose to define bias as a direct causal effect that is unjustified.
  We propose to define the closely related concept of disparity as a direct or indirect causal effect that includes a bias.
  Our proposed definitions can be used to study biases and disparities in a more rigorous and systematic way.
  We compare our definitions of bias and disparity with various criteria of fairness introduced in the artificial intelligence literature.
  In addition, we discuss how our definitions relate to discrimination.
  We illustrate our definitions of bias and disparity in two case studies, focusing on gender bias in science and racial bias in police shootings.
  Our proposed definitions aim to contribute to a better appreciation of the causal intricacies of studies of biases and disparities.
  We hope that this will also promote an improved understanding of the policy implications of such studies.
\end{abstract}


\maketitle

\section{Introduction}

\noindent
Bias is a central concept in the social and behavioural sciences, appearing in thousands of publications in a wide variety of contexts.
However, despite its widespread use, the concept of bias is often employed in ambiguous or imprecise ways.
For example, in studies of gender bias, ethnic bias, racial bias or class bias, the concept of bias often lacks a clear definition.
To enable a more precise discussion of biases, we propose an explicit definition rooted in structural causal models.
We also provide a definition of the closely related concept of disparity.

The typical statistical definition of bias is the difference between the expected estimated value of a parameter and the `true' value of the parameter \citep{Abramovich2013-he}.
A well-known example is the estimation of the variance of a distribution, where the simple sample variance is biased, and a small correction must be made to obtain an unbiased estimator.
Another example is selection bias, such as a sample of survey respondents that is not representative of the population of interest.

The concept of bias is also frequently used in the field of psychology, which has considered numerous cognitive biases \citep{Kahneman2011-wk}.
For example, it has been found that the decisions people make in monetary bets do not conform to rational choice utility models.
People tend to prefer less risky bets over more risky bets \citep{Kahneman2012-jd}.
Because this risk aversion differs from the theoretical outcome of rational choice utility models, risk aversion is often seen as a bias.
Whereas statistical biases refer to deviations of an estimator from the `true' value of a parameter, cognitive biases refer to deviations of human behaviour from a theoretical model.
However, whether human behaviour suffers from cognitive biases depends on the theoretical model to which human behaviour is compared.
For example, risk aversion might represent a cognitive bias compared to a rational choice utility model, but perhaps not compared to a prospect theory model \citep{Kahneman2012-jd}.

Implicit or unconscious bias \citep{Greenwald1995-cs} is another extensively studied concept of bias studied in the field of psychology.
People may have conscious and explicit ideas and attitudes and express them openly.
For example, a person might explicitly state that they prefer an apple over a pear or ABBA over The Prodigy.
People can also hold such ideas and attitudes implicitly, that is, without being consciously aware of them.
For example, a person might be inclined to pick an apple instead of a pear from a fruit basket, even if they claim to have no preference for apples over pears.
Likewise, when a researcher chooses between reading either an article authored by a man colleague or a similar article authored by a woman colleague, the researcher may be more likely to choose the former article, even if they claim to have no preference for articles authored by men.
An implicit preference for an apple over a pear will not typically be considered an implicit bias because there is no normative ideal that apples and pears should be equally coveted.
In contrast, an implicit preference for an article authored by a man over an article authored by a woman would presumably be considered an implicit bias because it violates the normative ideal of treating men and women equally.    
Instead of focusing on deviations from a `true' value or a theoretical model, the notion of implicit bias focuses on implicit preferences that deviate from a normative ideal.
The focus on a normative ideal is a key point: implicit bias concerns injustice or unfairness.

Many studies use the concept of bias without unambiguously clarifying how the concept is understood.
For example, when a study finds a difference between men and women, it is often presented as a gender bias.
Suppose a study shows that on average women perform better in poker games than men and suppose that the study presents this as a gender bias.
What does it mean to say there is a gender bias in poker games?
Should bias be interpreted as an inaccurate estimation of the `true' performance of men and women poker players?
Should bias be understood as behaviour that deviates from a theoretical model of rational choice and optimal poker play?
Or should bias be seen as an implicit attitude of poker players, treating women or men unfairly?
Without an unambiguous definition of bias, it is unclear what it means to claim that there is a gender bias in poker games.

Although researchers are often not explicit about this, they usually seem to understand the concept of bias in terms of causality.
For example, in observational studies of gender bias or racial bias, researchers usually control for confounding factors.
In the example about poker, researchers might control for the number of years of poker-playing experience and perhaps also for memory skills.
In doing so, researchers seemingly attempt to identify the direct causal effect of gender on poker performance: gender differences in poker performance do not result merely from women having more poker-playing experience; instead, these differences reflect the direct causal effect of gender on poker performance.

In this paper, we propose a definition of bias that captures two of the key ideas mentioned: (1) a bias represents a direct causal effect of one variable on another; (2) a bias represents an effect that is considered unjustified.
We use the framework of structural causal models introduced by \citet{Pearl2009-vd} to define bias in an unambiguous way.

We complement our definition of bias with a definition of the closely related concept of disparity.
We understand disparity as a broader concept than bias.
We define a disparity as a direct or indirect causal effect that includes a bias.

This paper is organised as follows.
In Section~\ref{sec:definitions}, we discuss structural causal models and use these models to introduce our definitions of bias and disparity.
In Section~\ref{sec:causal_inference}, we use structural causal models to highlight some challenges in the study of biases and disparities.
Bias and fairness play an important role in recent debates in the field of artificial intelligence (AI) \citep{ONeil2016-sa,Fry2019-ry}, leading to many suggestions for formal definitions of fairness in AI \citep{Oneto2020-fh}.
In Section~\ref{sec:fairness}, we compare our definitions of bias and disparity to some of the fairness criteria introduced in the AI literature.
We discuss how our definitions of bias and disparity may affect policy interventions in Section~\ref{sec:policy}.
In Section~\ref{sec:case_studies}, we illustrate our definitions of bias and disparity in two case studies.
The first case study concerns gender bias in science (Section~\ref{subsec:gender_bias}).
The second case study addresses racial bias in police shootings (Section~\ref{subsec:police_shooting}).
Finally, we offer some concluding remarks in Section~\ref{sec:discussion}.

\section{Defining bias and disparity}
\label{sec:definitions}

\noindent
We first briefly introduce the framework of structural causal models, which provides the foundation upon which we build our definitions of bias and disparity.
We refer to \citet{Pearl2009-vd} for a more in-depth treatment of structural causal models.
A more accessible introduction is available in \citet{Pearl2016-jx}, while \citet{Pearl2018-qo} provides a popular science account.
We aim to keep the introduction as simple as possible, covering only those elements of structural causal models that are essential for our definitions of bias and disparity.

\begin{figure}
  \includegraphics{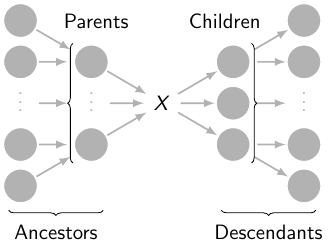}
  \caption{Illustration of our terminology.}
  \label{fig:terminology}
\end{figure}

\subsection{Structural causal models}

\noindent
Structural causal models are based on directed acyclic graphs (DAGs) which represent causal relationships between variables.
Each node in a DAG represents a variable, and we use the terms `node' and `variable' interchangeably.
We denote a link from node $X$ to node $Y$ by $X \rightarrow Y$.
Such a link represents a direct causal effect of $X$ on $Y$.
A DAG must not contain directed cycles.
For example, if a DAG contains a link $X \rightarrow Y$ and a link $Y \rightarrow Z$, it cannot contain a link $Z \rightarrow X$, but it may contain a link $X \rightarrow Z$.

If there is a directed path from node $X$ to node $Y$, then $X$ causally affects $Y$.
That is, if $X$ had been different, $Y$ would also have been different.
Causality has a direction: if $X \rightarrow Y$, then $X$ causally affects $Y$, but $Y$ does not causally affect $X$.
That is, if $Y$ had been different, this would not have affected $X$ because $Y$ is a result of $X$, and $X$ is not a result of $Y$.

Nodes that are directly affected by some node $X$ are called children of $X$, and nodes that directly affect $X$ are referred to as parents of $X$.
For example, if $X \rightarrow Y$, we call $X$ a parent of $Y$ and $Y$ a child of $X$.
Children, children of children and any nodes further downstream are called descendants.
Similarly, parents, parents of parents and any nodes further upstream are called ancestors (Fig.~\ref{fig:terminology}).
Hence, parents causally affect their children directly.
Ancestors causally affect their descendants either directly or indirectly.

A DAG offers a model for describing causal relationships and for systematic theoretical reasoning about such relationships.
Whether the causal relationships described by a DAG match empirical observations is a separate question.
Similar to most models, a DAG is typically unable to describe the real world in a fully accurate and comprehensive way.
Nonetheless, a DAG can be a useful tool for describing the most relevant causal relationships in a particular context.
In some cases, a DAG can be shown to be incompatible with empirical observations.
The DAG should then be rejected as a representation of the real world.
If a DAG is compatible with empirical observations, it may be considered a useful simplified description of the real world, at least tentatively.
However, it is important to realise that, typically, multiple DAGs are compatible with a particular set of empirical observations.
This reflects the existence of competing theoretical models of the real world.

Structural causal models offer the language to provide precise definitions of bias and disparity.
We are interested in bias and disparity in terms of some variable of interest $X$ and some outcome $Y$.
For example, when we speak of gender bias in citations, $X$ refers to gender and $Y$ to citations.
Likewise, when we speak of racial disparity in school acceptance, $X$ refers to race and $Y$ refers to school acceptance.
Often, but not always, the variable of interest $X$ represents a personal characteristic, such as gender, race or religion, which legal texts sometimes refer to as `protected characteristics' because they are protected under anti-discrimination law.
We do not use the term `protected characteristic' because the concepts of bias and disparity are not limited to protected characteristics, and the variable of interest $X$ may also represent other characteristics.
For example, \citet{Lee2013-rb} discuss prestige bias in peer review, where peer review is influenced by the prestige of a researcher or a research organisation.
In this case, the variable of interest is prestige.
As another example, \citet{Wang2017-ap} discuss novelty bias in publishing, where more novel research is less likely to be published in high-impact journals.
In this case, the variable of interest is novelty.

\begin{figure}
  \includegraphics{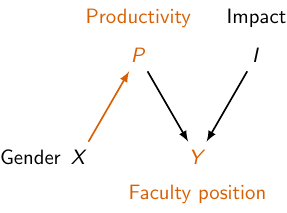}
  \caption{Simple hypothetical example illustrating our definitions of bias and disparity.
  Gender has a direct causal effect on productivity.
  This effect is considered unjustified and is therefore coloured red.
  We say there is a gender bias in productivity.
  Productivity and impact both have direct causal effects on faculty position.
  These effects are considered justified.
  The unjustified effect of gender on productivity affects faculty position indirectly.
  Therefore, both productivity and faculty position are considered unfair outcomes of gender and, hence, coloured red.
  We say there is a gender disparity in productivity and faculty position.
  }
  \label{fig:definitions}
\end{figure}

Fig.~\ref{fig:definitions} provides a simple hypothetical example of a structural causal model.
We will use this example to illustrate our definitions of bias and disparity.
In this example, the gender of a researcher affects the researcher's publication productivity.
Productivity, in turn, affects faculty position.
Faculty position is also affected by citation impact.

\subsection{Defining bias}

\noindent
We define bias as a direct causal effect that is considered unjustified.
If there is a direct causal effect of $X$ on $Y$, and this effect is considered unjustified, it constitutes a bias.
We refer to this as a bias of $X$ in $Y$, denoted by $X \rightbias Y$, where we use the red arrow to indicate that the direct causal effect of $X$ on $Y$ is considered unjustified.
Whether a particular direct causal effect is justified is an ethical question that cannot be determined empirically using data.
A bias may be implicit or explicit.
This distinction plays no role in our definition.

In the example presented in Fig.~\ref{fig:definitions}, there is a gender bias in productivity because gender has a direct causal effect on productivity, and this effect is considered unjustified.
Productivity and impact both affect faculty position.
Because these effects are considered justified, they do not represent biases.

\subsection{Defining disparity}

\noindent
We define a disparity as a causal effect of $X$ on $Y$ that includes a bias.
We refer to this as a disparity of $X$ in $Y$.
There is a disparity of $X$ in $Y$ if at least one link on a causal pathway from $X$ to $Y$ represents a bias.
More formally, there is a disparity of $X$ in $Y$ if there exists a directed path $X \rightarrow \ldots X' \rightbias Y' \rightarrow \ldots \rightarrow Y$, where $X' \rightbias Y'$ denotes a bias of $X'$ in $Y'$.
Disparity is a broader concept than bias.
Each bias is a disparity, but a disparity does not need to be a bias.
If there is a disparity of $X$ in $Y$, we consider the outcome $Y$ to be unfair with respect to $X$.

A bias $X \rightbias Y$ renders both $Y$ and all descendants of $Y$ unfair.
This is illustrated by the example presented in Fig.~\ref{fig:definitions}.
There is no gender bias in faculty position in this example because there is no direct causal effect of gender on faculty position.
However, there is an indirect causal effect: gender affects productivity and productivity, in turn, affects faculty position.
This indirect causal effect produces a gender disparity in faculty position.
This gender disparity is due to the gender bias in productivity.
Hence, although each direct causal effect on faculty position is considered justified, faculty position is an unfair outcome of gender.
This illustrates how a single bias renders many outcomes unfair.

As another example, consider the seminal case of  Griggs v.\ Duke Power Co., 401 U.S.\ 424, 430--31 (1971)\footnote{\url{https://supreme.justia.com/cases/federal/us/401/424/}}.
The Supreme Court ruled that the requirement of a high-school diploma was not justified by the business needs of Duke Power Co.
This unjustified requirement disadvantaged applicants of some races because race affected the probability of obtaining a high-school diploma.
In our terminology, there is no racial bias in the hiring practices of Duke Power Co.
Instead, there is a diploma bias in the company's hiring practices that produces a racial disparity in hiring.

As a third example, consider the practice of redlining in the United States, in which organisations such as insurers deny people services based on the area in which they live.
In practice, due to racial segregation, this amounts to selectively not serving people of a certain race.
In our terminology, there could be multiple biases.
Using ZIP codes to determine whom to insure may be considered unjustified, in which case there is a location bias in insuring and a racial disparity in insuring.
There could also be a racial bias in neighbourhoods if people of a certain race are denied access to certain neighbourhoods.
This racial bias in neighbourhoods consequently produces a racial disparity in insuring, even if using ZIP codes were deemed justified.
If insurers use race to determine whom to insure, there is a racial bias in insuring, not only a racial disparity.
In this example, even if there is no racial bias in insuring, it would not imply that there is no problem.
A racial disparity in insurance indicates that the outcome is unfair with respect to race, which signals a problem.

\subsection{Discrimination}
\label{subsec:discrimination}

In addition to bias, discrimination is an often used concept in the social sciences.
Similar to bias, discrimination can cover various issues, such as age, gender or religious discrimination.
Especially the study of racial discrimination has a long history in the social sciences \citep{winant_race_2000}.
In contrast to bias, more formal definitions for (racial) discrimination have been discussed more extensively in the literature.
A seminal report in this context is ``Measuring Racial Discrimination'' by the \citet{national_research_council_measuring_2004} of the US.
The report provides an extensive overview of different approaches to measuring discrimination and challenges that surface when trying to do so, including causal challenges.
The report defines racial discrimination as follows:

\begin{quote}
  (1) differential treatment on the basis of race that disadvantages a racial group and
  (2) treatment on the basis of inadequately justified factors other than race that disadvantages a racial group (differential effect).
  \citep[p. 39]{national_research_council_measuring_2004}
\end{quote}

This definition in two components is common \citep{pager_sociology_2008}.
The first component is similar, but not identical, to our definition of bias (applied specifically to race), while the second component is similar, but not identical, to our definition of disparity (applied specifically to race).
The report does not distinguish between the two components and labels both components as discrimination.
This sometimes leads to confusing discussions in the report.
For example, many experimental approaches that the report discusses aim to identify a direct causal effect, or a bias, in our terminology.
However, when discussing observational approaches, it is not always clear whether the report discusses direct causal effects or total causal effects, obfuscating the discussion.
\citet{national_research_council_measuring_2004} also distinguishes between ``intentional'' and  ``subtle'' discrimination.
We do not make such a distinction in our definition of bias.

The literature also uses other terms related to discrimination, such as ``structural discrimination'', ``systemic discrimination'' and ``institutional discrimination'' \citep{small_sociological_2020}.
There does not seem to be agreement on the definitions of these terms: ``The term structural discrimination has been used loosely in the literature'' \citep[p. 17]{pager_sociology_2008} and ``these terms are not used consistently across the social sciences'' \citep[p. 52]{small_sociological_2020}.
One common idea seems to be that discrimination need not result from intentional actions driven by prejudice, but that it can also be a result of the application of (seemingly neutral) rules that lead to differential outcomes.
This is perhaps related to ``cumulative discrimination'' \citep{pager_sociology_2008}, which refers to downstream effects of discrimination.
Indeed, racial differences in the US are visible throughout various domains, including employment, housing and credit markets \citep{pager_sociology_2008}, and discrimination in one domain may affect outcomes in other domains, which is related to the idea of a ``discrimination system'' \citep{reskin_race_2012}.
We capture such downstream effects in our definition of disparity.
While our definitions of bias and disparity do not capture all ideas discussed in the literature on discrimination, we believe that our distinction between bias and disparity can help to bring greater clarity to the literature.

In economics, there is a tradition of distinguishing ``statistical discrimination'' from ``taste-based discrimination'' \citep{charles_studying_2011}.
Statistical discrimination refers to the use of group means to predict something about an individual, which might then be used in later decisions.
The traditional example is an employer who cannot directly observe the productivity of a job applicant, and uses a group average to predict an applicant's productivity to make a hiring decision.
In this situation, belonging to a certain group (e.g. a certain race) does not determine a hiring decision directly, but it does influence the decision indirectly.
Taste-based discrimination refers to having a preference for certain groups, which seems similar to ``intentional discrimination''.
One could argue that statistical discrimination refers to an indirect effect on the hiring decision, mediated by the prediction of productivity, while taste-based discrimination refers to a direct effect.
We will get back to the challenge of distinguishing a direct from an indirect effect in the next section.

\subsection{Causal concerns}
\label{subsec:causal_inference}

\noindent
In this section, we aim to draw attention to a few complex issues related to our causal definitions of bias and disparity.

First, the presence of a causal effect of a variable of interest does not imply that the variable of interest is also accountable for that effect~\citep{VanderWeele2014-zd}.
This requires careful attention, especially when studying variables such as gender, race and religion.
Many decisions involve both someone who decides and someone for whom something is decided.
For example, in funding applications, decisions are made by a review panel rather than the applicants.
The review panel should be held accountable for any bias in funding decisions.
The applicants should not be held accountable.

Second, it is important to acknowledge that causal effects may be due to prevailing societal roles, norms or contexts.
For example, certain cultural preferences, such as women preferring pink, are socially constructed and have changed over time \citep{Grisard2017-pn}.
Understanding such cultural dynamics is an important research topic in the social sciences.
Culturally and socially constructed roles and patterns may shape causal effects.
Given certain cultural and societal roles and patterns, we can identify causal effects of gender, ethnicity or race, but this does not mean that these causal effects will remain unchanged over time.
On a long enough timescale, such roles and patterns can change, while on a shorter timescale, they structure much of the social interaction.

Third, when a variable of interest is a personal characteristic, such as gender or race, it is sometimes contested whether the variable can be considered a `causal factor'.
Indeed, \citet{Holland2003-wh} argues that race should not be understood as having a causal effect, summarised in the popular aphorism, `no causation without manipulation'~\citep{Holland1986-ol}.
Gender and race are frequently considered to have causal effects, and people reason about them as such.
Defining the concepts of bias and disparity without resorting to causality contrives the matter instead of explicating it.
Rather than considering whether these variables are manipulable, we can simply consider hypothetical possibilities~\citep{Pearl2018-qu}: what if a White person had been Black, or a man had been a woman?
Of course, this raises the difficulty of defining White or Black or conceptualising what it entails to be a woman instead of a man, as discussed by \citet{Kohler-Hausmann2018-at}.
Should clothing change when we consider the hypothetical possibility of someone being a man instead of a woman?
Should the hairstyle change?
Should jewellery be changed?
Indeed, a man in women's clothing could produce a different effect than a man in men's clothing.

These problems are not limited to personal characteristics.
Consider, for example, the interdisciplinarity of a paper, which was reported to show a bias in citations \citep{Rinia2001-yx}.
How should we conceptualise interdisciplinarity, and how should we interpret hypothetical possibilities?
Does the topic change alongside the paper's interdisciplinarity?
The writing style?
The analysis performed?
How can we hypothetically change the interdisciplinarity of a paper without also changing other aspects?

These conceptual issues are indeed challenging, and we cannot provide definitive answers here.
However, this does not mean that we should discard research into gender, racial and interdisciplinarity biases.
Instead, we should endeavour to carefully and precisely address these challenging conceptual issues.

\begin{figure}
  \includegraphics{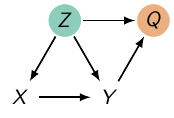}
  \caption{Example of a simple DAG.
  On a path between nodes $X$ and $Y$, node $Z$ is a confounder (hence, open), and node $Q$ is a collider (hence, closed).}
  \label{fig:example_simple}
\end{figure}

\section{Difference does not imply bias or disparity}
\label{sec:causal_inference}

\noindent
Most researchers will be familiar with the adage `correlation does not imply causation'.
A key question researchers often face is whether a correlation between two variables represents a causal effect.
A similar question arises when we observe a difference in outcomes: does the difference represent a bias or disparity?
If the difference represents a causal effect, there may be a disparity or even a bias.
If the difference does not represent a causal effect, there is no bias or disparity.
For example, we may observe a gender difference in citations.
This raises several questions: does this difference represent a (direct or indirect) causal effect of gender on citations?
Do publications of authors of a particular gender receive fewer citations because of the gender of the authors?
If there is indeed such a causal effect of gender on citations, most readers would probably agree that the effect is unjustified.
The effect then constitutes a gender disparity in citations and perhaps even a gender bias.
If there is no causal effect, there is no gender bias or gender disparity in citations.

Structural causal models offer a useful tool to understand whether a difference does or does not represent a causal effect.
Two variables may be associated without being causally related.
The following subsections explain in basic terms how to determine whether two variables are associated and whether such an association does or does not represent a causal relationship.
This is critical for understanding whether a difference may represent a bias, a disparity or neither.

\subsection{\texorpdfstring{$d$}{d}-connectedness and \texorpdfstring{$d$}{d}-separation}

\begin{figure*}
  \includegraphics{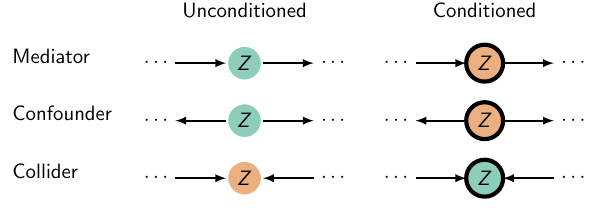}
  \caption{
    Illustration showing when a node $Z$ on an undirected path is open (coloured green) or closed (coloured red).
    Conditioning on a variable `flips' a node from open to closed or vice versa.
  }
  \label{fig:summary}
\end{figure*}

\noindent
The concepts of $d$-connectedness and $d$-separation indicate whether two variables in a DAG are respectively associated or independent.
If two variables are $d$-connected, they are associated.
If they are $d$-separated, they are independent.
The concepts of $d$-connectedness and $d$-separation may be somewhat difficult to comprehend.
We explain these concepts briefly below.
For a more extensive introduction to $d$-connectedness and $d$-separation, we refer to \citet{Pearl2009-vd}, especially Chapter~11.1.2, aptly titled `$d$-separation without tears'.

First, we introduce the concept of open and closed undirected paths between two nodes in a DAG.
An undirected path comprises a sequence of nodes connected via links that may point in either direction.
For example, $X \rightarrow Z \rightarrow Y$ is an undirected path, and so are $X \leftarrow Z \rightarrow Y$ and $X \rightarrow Z \leftarrow Y$.
An undirected path is open if all nodes on the path are open.
A node $Z$ on a path is open if it is connected as $\ldots \rightarrow \innode{Z}{open} \rightarrow \ldots$, called a mediator, or if it is connected as $\ldots \leftarrow \innode{Z}{open} \rightarrow \ldots$, called a confounder.
A node $Z$ is closed if it is connected as $\ldots \rightarrow \innode{Z}{closed} \leftarrow \ldots$, called a collider.
We indicate whether a node is open or closed by the colour of the node, where $\innode{Z}{open}$ represents an open node and $\innode{Z}{closed}$ represents a closed node (Fig.~\ref{fig:example_simple}).
A node may play different roles on different paths.
For example, the same node may act as a mediator on one undirected path, as a confounder on another undirected path and as a collider on yet another undirected path.
Colliders play an important role in some of this paper's discussions.
In short, undirected paths without any colliders are open, and undirected paths with one or more colliders are closed.
Instead of open or closed undirected paths, we simply refer to these as open or closed paths.

In a sense, open paths allow information to flow freely between variables, while closed paths somehow block the flow of information.
If nodes $X$ and $Y$ are connected by at least one open path, the nodes are $d$-connected, and information can flow freely between them.
If there are no open paths between $X$ and $Y$, the nodes are $d$-separated, and no information can flow between them.
Two variables $X$ and $Y$ that are $d$-connected are associated.
That is, if $X$ and $Y$ are $d$-connected, observing $X$ tells you something about $Y$ and vice versa.
Two variables $X$ and $Y$ that are $d$-separated are independent: observing $X$ tells you nothing about $Y$.
The association between two variables that are $d$-connected does not need to reflect causality.
The simplest example is $X \leftarrow \innode{Z}{open} \rightarrow Y$, where the confounder $Z$ affects both $X$ and $Y$, so that $X$ and $Y$ are correlated only because of the common factor $Z$.
In contrast, if $X \rightarrow \innode{Z}{open} \rightarrow Y$, the variable $Z$ acts as a mediator and the association between $X$ and $Y$ does reflect causality.

Open and closed paths are sometimes referred to as unblocked and blocked paths.
Independence between two variables $X$ and $Y$ is sometimes denoted by $X \perp Y$.
Hence, if $X$ and $Y$ are $d$-separated, this can be denoted by $X \perp Y$.
If $X$ and $Y$ are associated, they are not independent, which is sometimes denoted by $X \not\perp Y$.
Hence, if $X$ and $Y$ are $d$-connected, this can be denoted by $X \not\perp Y$.
In summary, two variables are $d$-connected if there is at least one path between them with only confounders and mediators (Fig.~\ref{fig:summary}).
However, there is an important twist to $d$-connectedness and $d$-separation, which we discuss in the next subsection.

\subsubsection*{Conditioning and selection}

\noindent
Many studies condition on certain variables.
For example, studies frequently control for certain variables by including them in a regression analysis, which amounts to conditioning on these variables.
Some studies include only certain people in an analysis; for example, considering only people who have been arrested or only scholars with at least five publications.
Such selection criteria also amount to conditioning on a variable.
Other studies perform analyses on separate subsets of the data.
A common example in scientific studies is analysing different scientific fields separately.
Performing analyses on separate subsets of the data amounts to conditioning on the variables used to define the subsets, such as scientific field.
In other cases, scientific fields are not analysed separately, but instead certain variables, such as the number of citations of a publication, are field-normalised.
Normalising by a variable also amounts to conditioning on that variable.
Hence, conditioning on variables is a common occurrence, and it has profound implications for the notions of $d$-connectedness and $d$-separation.

When conditioning on a node, the node will become closed if it was open before conditioning and conversely, the node will become open if it was closed before conditioning (Fig.~\ref{fig:summary}).
That is, the open or closed status of a node inverts when conditioning on that node.
Hence, when conditioning on a node $Z$ on a path, the node is closed if it is connected as a mediator ($\ldots \rightarrow \innode{Z}{closed,conditioned} \rightarrow \ldots$) or a confounder ($\ldots \leftarrow \innode{Z}{closed,conditioned} \rightarrow \ldots$) and open if it is connected as a collider ($\ldots \rightarrow \innode{Z}{open,conditioned} \leftarrow \ldots$).
We represent conditioning using a thick enclosing circle $\innode{Z}{conditioned}$.
Hence, if a path is open, it can be closed by conditioning on a mediator or confounder on the path.
A path that is closed can be opened by conditioning on a collider.%
  \footnote{There is an additional complication: a collider becomes open not only by conditioning on the collider itself but also by conditioning on any of its descendants.
    For example, if $X \rightarrow Z \leftarrow Y$ and also $Z \rightarrow W$, conditioning on $W$ (partially) opens $Z$.
    We will not discuss this further.
    For more information, see \citet[Chapter~11.1.2]{Pearl2009-vd}.
    }%

Because a node may act as a confounder or mediator on one path and as a collider on another path, conditioning on a node may close one path but open another one.
In Fig.~\ref{fig:example_collider}a, the path $X \rightarrow \innode{Z}{open} \rightarrow Y$ is open because $Z$ acts as a mediator on this path, and the path $X \rightarrow \innode{Z}{closed} \leftarrow \innode{U}{open} \rightarrow Y$ is closed because $Z$ acts as a collider on this path.
If we condition on $Z$, we close the path $X \rightarrow \innode{Z}{closed,conditioned} \rightarrow Y$, where $Z$ acts as a mediator, and we open the path $X \rightarrow \innode{Z}{open,conditioned} \leftarrow \innode{U}{open} \rightarrow Y$, where $Z$ acts as a collider.

If $X$ and $Y$ are $d$-separated when conditioning on $Z$, they are said to be conditionally independent.
This is sometimes denoted by $X \perp Y \mid Z$.
If $X$ and $Y$ are $d$-connected when conditioning on $Z$, they are not independent, which can be denoted by $X \not\perp Y \mid Z$.

\subsection{Challenges in identifying biases and disparities}

\noindent
A causal effect can be challenging to identify, complicating the identification of a bias and a disparity.
Researchers often approach this problem by taking the `causal salad' approach \citep{McElreath2020-wj}: include every possibly relevant factor in the hope of obtaining a close approximation of a causal effect.
In essence, the following reasoning is used: controlling for $A$, $B$, $C$ and $D$, we still see a difference in $Y$ based on $X$ and therefore it is highly likely that $X$ has a causal effect on $Y$.
In studies on discrimination, researchers name this difference after controlling for all these factors the ``residual race gap'' \citep{pager_sociology_2008}.
However, the presence of colliders is at odds with this approach: controlling for colliders \emph{prevents} the identification of a causal effect.
If colliders are simultaneously also mediators or confounders, the problem is even worse, especially if there are some unobserved or unobservable variables.
To identify a causal effect, we must ensure that all non-causal paths are closed, and that only the relevant paths are open.

\begin{figure}
  \includegraphics{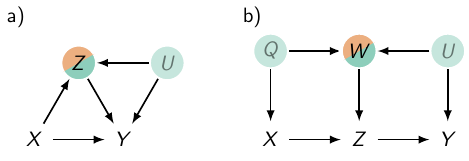}
  \caption{Challenges in identifying biases and disparities.
  The variables $U$ and $Q$ are assumed to be unobserved or unobservable, making it impossible to control for them.}
  \label{fig:example_collider}
\end{figure}

Consider the example provided in Fig.~\ref{fig:example_collider}a.
Suppose that we are interested in identifying a bias of $X$ in $Y$.
Because we are interested in the direct causal effect of $X$ on $Y$, we need to control for the mediator $Z$, closing the path $X \rightarrow \innode{Z}{closed,conditioned} \rightarrow Y$.
However, $Z$ also acts as a collider on the path $X \rightarrow \innode{Z}{open,conditioned} \leftarrow \innode{U}{open,unobserved} \rightarrow Y$, and conditioning on $Z$ opens this non-causal path.
This poses a conundrum for identifying the direct causal effect of $X$ on $Y$: if we condition on $Z$, we condition on a collider, but if we do not condition on $Z$, we fail to control for a mediating effect.
In fact, if $U$ is unobserved, there is no straightforward way of identifying the direct causal effect of $X$ on $Y$.

Identifying a disparity can be equally challenging in the presence of colliders.
Consider the example provided in Fig.~\ref{fig:example_collider}b.
Suppose that we are interested in identifying a disparity of $X$ in $Y$.
If we consider the effect $X \rightarrow Z$ or the effect $Z \rightarrow Y$ as unjustified, the total causal effect of $X$ on $Y$ is a disparity of $X$ in $Y$.
The path $X \leftarrow \innode{Q}{open,unobserved} \rightarrow \innode{W}{open} \rightarrow \innode{Z}{open} \rightarrow Y$ is open but does not represent a causal effect of $X$ on $Y$.
We need to close this path to identify the causal effect of $X$ on $Y$.
We cannot condition on $Z$ to close this path because we are interested in the total causal effect, and conditioning on $Z$ also closes the causal path $X \rightarrow \innode{Z}{closed,conditioned} \rightarrow Y$.
If we condition on $W$, we close the non-causal path.
However, $W$ also acts as a collider on the path $X \leftarrow \innode{Q}{open,unobserved} \rightarrow \innode{W}{open,conditioned} \leftarrow \innode{U}{open,unobserved} \rightarrow Y$, and conditioning on $W$ opens this non-causal path.
This poses a conundrum for identifying the total causal effect of $X$ on $Y$: if we condition on $W$, we condition on a collider, but if we do not condition on $W$, we fail to control for a confounding effect.
If $Q$ and $U$ are unobserved, there is no straightforward way of identifying the total causal effect of $X$ on $Y$.

Unlike the estimation of a disparity, the estimation of a bias is sensitive to refinements of causal pathways.
For example, suppose that we initially assume that a structural causal model consist only of $X \rightbias Y$, which is considered a bias.
Let us call this model $A$.
Additionally, suppose that, after further consideration, there turns out to be a mediating factor $Z$, resulting in $X \rightbias \innode{Z}{open} \rightbias Y$, where the effects $X \rightbias Z$ and $Z \rightbias Y$ are both considered biases.
Let us call this model $B$.
The additional mediating factor $\innode{Z}{open}$ in model $B$ does not change the total causal effect of $X$ on $Y$.
This means that the disparity of $X$ in $Y$ is the same in models $A$ and $B$.
However, there is a difference for the direct causal effect of $X$ on $Y$.
In model $A$, we do not need to control for anything, and we can estimate the direct causal effect directly from observing $X$ and $Y$, showing that there is a bias of $X$ in $Y$.
In contrast, in model $B$, we need to control for the mediating factor $\innode{Z}{conditioned,closed}$ to estimate the direct causal effect of $X$ on $Y$.
Doing so shows that there is no bias of $X$ in $Y$.

This example illustrates how refinements of causal pathways challenge claims of biases.
An effect might be not direct but mediated via variables not initially included in the causal model.
This connects to the discussion in Section~\ref{subsec:causal_inference} about the interpretation of hypothetical possibilities such as: what if a man were a woman, and
if so, would their clothing have been different as well?
If the concept of gender comprises clothing, differences in clothing would be part of a direct effect of gender.
Otherwise, clothing might be a separate, mediating, variable for gender.
Indeed, discussions about biases may involve suggestions for more refined causal pathways or questions about what exactly a specific hypothetical change would entail.
This issue does not play a role in analyses of disparities: total causal effects are unaffected by refinements of causal pathways.

\section{Fairness in AI}
\label{sec:fairness}

\begin{figure*}
  \includegraphics{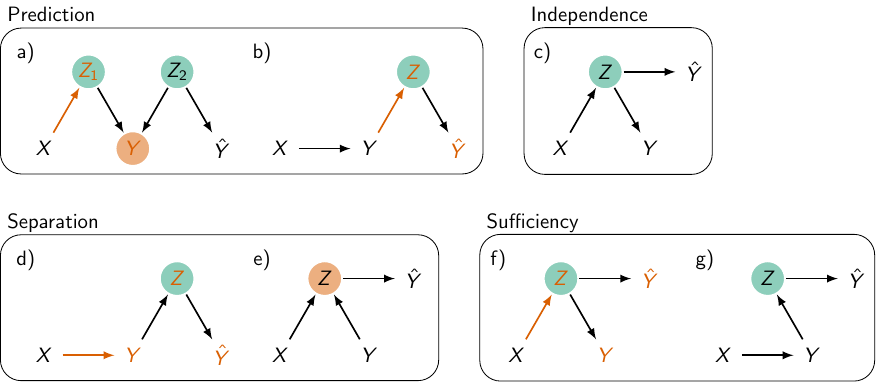}
  \caption{Illustration of various fairness criteria in AI and how they are at odds with the notion of fairness underlying our definitions of bias and disparity.
  (a) Although there is a disparity in $Y$, there is no disparity in the prediction $\hat{Y}$.
  (b) Conversely, although there is no disparity in $Y$, there is a disparity in the prediction $\hat{Y}$.
  (c) Although there is no disparity in $\hat{Y}$, the prediction is considered unfair according to the independence criterion.
  (d) Although there is a disparity in $\hat{Y}$, the prediction is considered fair according to the separation criterion.
  (e) Conversely, although there is no disparity in $\hat{Y}$, the prediction is considered unfair according to the separation criterion.
  (f) Although there is a disparity in $\hat{Y}$, the prediction is considered fair according to the sufficiency criterion.
  (g) Conversely, although there is no disparity in $\hat{Y}$, the prediction is considered unfair according to the sufficiency criterion.
  }
  \label{fig:fairness}
\end{figure*}

\noindent
Related discussions are taking place in AI research, where fairness is an increasingly important consideration.
There are several good overviews of fairness in AI \citep{Mehrabi2019-jg,Barocas2020-ui}.
We will follow the overview provided by \citet{Barocas2020-ui} because their presentation is closely aligned with a causal framework.
Much of the work in this area focuses on data-driven criteria to decide whether certain predictions should be considered fair.
Most work is not concerned with the fairness of outcomes, but is limited to the fairness of predictions only.
We show that our definitions of bias and disparity can also be applied to predictions instead of outcomes.
We discuss three popular fairness criteria proposed in the literature---independence, separation and sufficiency---and show that these criteria are uninformative compared to our concepts of bias and disparity.

\subsection{Prediction}

\noindent
The focus in AI is typically on predicting a certain outcome based on certain predictors.
A central question is whether the prediction is fair.
Our focus is different: we are concerned with whether a particular outcome is fair.
Nonetheless, there are connections between the two questions, and we can also apply our concepts of bias and disparity to a prediction.
Suppose we try to predict the outcome $Y$ based on the features $\mathcal{Z} = \{Z_1,\ldots,Z_k\}$.
In the DAG representation, the predictors $\mathcal{Z} = \{Z_1, \ldots, Z_k\}$ are the parents of the prediction $\hat{Y}$.
The typical question is whether $\hat{Y}$ is fair for some variable of interest $X$.
The various fairness criteria all try to answer this question.
Our concepts of bias and disparity can also be applied to $\hat{Y}$.
We consider $\hat{Y}$ unfair for some variable of interest $X$ if there is a disparity of $X$ in $\hat{Y}$, and fair if there is no such disparity.

The fairness of $\hat{Y}$ is independent of the fairness of $Y$; the one can be fair while the other can be unfair.
For example, as Fig.~\ref{fig:fairness}a illustrates, suppose there is a social class bias in schooling level, $X \rightbias Z_1$, which, in turn, affects job prospects $Y$, which are also affected by work ethic $Z_2$.
Then, if we predict job prospects $\hat{Y}$ based only on work ethic $Z_2$, the prediction of job prospects shows no social class disparity and should be considered fair, even though the actual job prospects $Y$ do show a social class disparity and should be considered unfair.

Conversely, consider the example illustrated in Fig.~\ref{fig:fairness}b.
Suppose that gender $X$ affects arrogance $Y$ and that there is a bias $Y \rightbias Z$ of arrogant people being more frequently selected for leadership positions.
When predicting arrogance $\hat{Y}$ based on leadership position $Z$, there will be a gender disparity in $\hat{Y}$, even though there is no gender disparity in $Y$.
That is, the fairness of $\hat{Y}$ implies nothing about the fairness of $Y$.

\subsection{Independence criterion}

\noindent
The \emph{independence fairness criterion} holds if the variable of interest $X$ is independent of the prediction $\hat{Y}$, denoted by $X \perp \hat{Y}$.
This is sometimes known as demographic parity or statistical parity.
If $X \perp \hat{Y}$, there is clearly no causal effect of $X$ on $\hat{Y}$, and there can be no unfairness according to our definition.
However, although it might be intuitively appealing to demand complete independence, that would also forego any possibility of justified differences.
The independence criterion considers any influence of $X$ unfair, even if there is no disparity according to our definition.

For example, suppose gender $X$ affects thrill-seeking preference $Z$, which, in turn, affects whether a person goes bungee jumping $Y$, as Fig.~\ref{fig:fairness}c illustrates.
If we predict an interest in bungee jumping $\hat{Y}$ based on thrill-seeking preference $Z$, the prediction will not be independent of gender $X$.
The prediction is therefore considered unfair according to the independence criterion, but fair according to our definition because there is no gender disparity in $\hat{Y}$.

\subsection{Separation criterion}

\noindent
The \emph{separation fairness criterion} stipulates that $\hat{Y} \perp X \mid Y$.
This means that the prediction $\hat{Y}$ is independent of the variable of interest $X$ when we control for the actual outcome $Y$.
The separation criterion implies that the prediction shows the same error rate for each value of $X$, such as for each gender or race.
For this reason, it is also known as equalised odds.

Although the separation criterion may seem intuitively reasonable, it is completely contrary to our approach: equal error rates reproduce existing biases.
That is, if there is a disparity of $X$ in $Y$, equal error rates simply reproduce this disparity in the prediction $\hat{Y}$.
Correcting the disparity of $X$ in $Y$ actually requires \emph{different error rates} for different values of $X$.

If we predict $Y$ based only on descendants of $Y$ that are not otherwise descendants of $X$, then $\hat{Y}$ and $X$ are $d$-separated by $Y$, that is, $\hat{Y} \perp X \mid Y$.
Consequently, the separation criterion is satisfied, even though there is a disparity of $X$ in $\hat{Y}$.
For example, as Fig.~\ref{fig:fairness}d illustrates, suppose that there is a racial bias in job prospects, $X \rightbias Y$, and that having a job $Y$ affects your income $Z$.
If we now predict having a job $\hat{Y}$ based on income $Z$, the prediction $\hat{Y}$ is independent of race $X$, given the observation of having a job $Y$.
Therefore, $\hat{Y}$ satisfies the separation criterion.
However, according to our definition, $\hat{Y}$ shows a racial disparity.
Hence, the separation criterion may consider a prediction to be fair that is considered unfair according to our definition.

Suppose, on the contrary, that the separation criterion does not hold, so that $X \not\perp \hat{Y} \mid Y$.
Then $X$ and $\hat{Y}$ are $d$-connected, even when conditioning on $Y$.
This is a structural criterion that involves only conditional independencies and does not depend on any ethical judgement of whether an effect is justified.
That is, even if separation does not hold, this does not imply that $\hat{Y}$ is unfair according to our definition.
For example, as Fig.~\ref{fig:fairness}e illustrates, suppose race $X$ affects church attendance $Z$, which is also affected by religiosity $Y$.
If we predict religiosity $\hat{Y}$ based on church attendance $Z$, the prediction does not satisfy the separation criterion and is therefore considered unfair according to this criterion.
However, because there are no biases, the prediction is considered fair according to our definition.
This holds true in general: structural criteria cannot be used to determine whether a prediction is considered fair according to our definition.

\subsection{Sufficiency criterion}

\noindent
The \emph{sufficiency fairness criterion} stipulates that $Y \perp X \mid \hat{Y}$.
This means that the outcome $Y$ is independent of the variable of interest $X$ when we control for the prediction $\hat{Y}$.
The sufficiency criterion implies a parity of predictive values for each $X$, such as for each gender or race.
An AI model that satisfies the sufficiency criterion is sometimes said to be calibrated.

Although the sufficiency criterion may seem intuitively reasonable it is contrary to our approach.
If $\hat{Y}$ predicts sufficiently well the dependence between $X$ and $Y$, conditioning on $\hat{Y}$ will make $X$ and $Y$ (nearly) independent.
For example, as Fig.~\ref{fig:fairness}f illustrates, suppose there is a racial bias in income, $X \rightbias Z$, and income $Z$, in turn, affects credit card limits $Y$.
If we predict credit card limits $\hat{Y}$ based on income $Z$, then $\hat{Y}$ will essentially be a close proxy for $Z$.
This renders race conditionally independent of credit card limits, given the predicted credit card limits, $X \perp Y \mid \hat{Y}$, which means that the prediction is considered fair according to the sufficiency criterion.
However, there is a racial bias in income $Z$, so there is a racial disparity in predicted credit card limits $\hat{Y}$, and according to our definition, the prediction is therefore considered unfair.
Hence, according to our definition, the sufficiency criterion cannot distinguish between fair and unfair predictions.

Conversely, suppose that sufficiency does not hold.
Sufficiency is a structural criterion, and structural criteria cannot be used to determine whether a prediction $\hat{Y}$ is considered fair according to our definition.
For example, as Fig.~\ref{fig:fairness}g illustrates, suppose people who hold a different religion $X$ have a different musical taste $Y$, which produces a difference in whether they play the piano $Z$.
When predicting musical taste $\hat{Y}$ based on piano playing $Z$, the prediction does not satisfy sufficiency, but according to our definition, the prediction is fair.

\subsection{Causal fairness approaches}

\noindent
In our view, the discussed fairness criteria in AI cannot be used to determine whether outcomes or predictions of outcomes should be considered fair.
More recently, other approaches to fairness that consider causality have received substantial attention.
\citet{plecko_causal_2022} provide a comprehensive integrated overview of causal fairness approaches, and \cite{carey_causal_2022} discuss these approaches also in the broader philosophical, legal and sociological context.
The central challenge in fairness in AI is how a prediction $\hat{Y}$ can be made fair even if there is a disparity of $X$ in $Y$.
Causal approaches to fairness acknowledge that addressing this challenge requires a causal understanding.
In line with our approach, some of these approaches also acknowledge that the fairness of a causal effect is an ethical question that cannot be answered by structural criteria.
Our approach seems to align most closely with the counterfactual approach suggested by \citet{Kusner2017-cf}, \citet{Chiappa2019-yz} and \citet{Loftus2018-kd}, and continued by \citet{Oneto2020-fh}.
Overall, the counterfactual approach seems a viable and informative approach to fairness in AI, broadly in agreement with our proposed definitions of bias and disparity.

\section{Policy interventions}
\label{sec:policy}

\noindent
Understanding whether an observed difference represents a bias, a disparity or neither is important for making appropriate suggestions for policy interventions.
If a difference does not represent a bias or disparity, there is probably no need to intervene.
Whether a difference represents a bias or disparity depends on whether there is a causal effect and whether this effect is considered unjustified.
If there is no causal effect, there can be no bias or disparity.
For example, as previously discussed, sometimes a difference arises due to selection on a collider, in which case the difference does not represent a causal effect.
If such a difference is incorrectly interpreted as a bias or disparity, attempts to correct it are likely to have unexpected consequences.
We will encounter this issue in the case studies discussed in the next section.

If it is established that there is an unjustified causal effect that represents a disparity, this offers grounds for intervening to correct the injustice.
Importantly, the appropriate intervention depends on where the bias is located.
For example, if $X \rightbias \innode{Z}{open} \rightarrow Y$, there is a disparity of $X$ in $Y$ due to a bias of $X$ in $Z$.
There is also a causal effect of $Z$ on $Y$, but this effect is not deemed unjustified and therefore does not represent a bias.
If the bias is incorrectly perceived to be located between $Z$ and $Y$ instead of between $X$ and $Z$, an intervention in the process between $Z$ and $Y$ may be suggested.
However, because there is no bias of $Z$ in $Y$, such an intervention will not be effective and will not correct the disparity of $X$ in $Y$.

Interventions such as gender quotas or affirmative action may also be considered to address the disparity of $X$ in $Y$.
Although such interventions may indeed remove the difference of $X$ in $Y$, they do not correct the bias of $X$ in $Z$.
Instead of correcting this bias, a new causal effect of $X$ on $Y$ is added.
Depending on the context, this may have undesirable consequences.

For example, suppose there is a gender bias in childcare, $X \rightbias Z$, which in turn affects who is hired, $Z \rightarrow Y$: women perform more childcare than men, and those who perform more childcare are less likely to be hired.
In this example, a policy aimed at addressing gender biases in hiring will have no effect because there is no such bias.
To address the gender disparity in hiring, one might consider introducing a quota for women when hiring new employees.
This will lower the chances of men being hired, which might be perceived as justified because of the need to counter the gender disparity in hiring.
However, the quota will decrease the chance of being hired for men who take care of children and increase the chance of being hired for women who do not take care of children.
This decreases the proportion of men who take care of children in the working population and increases the proportion of women who do not take care of children in the working population.
Thus, the quota might be considered counterproductive, potentially reinforcing the idea that one should not take care of children if one wants a career.

In some situations, interventions that directly address a particular bias are not possible.
This might justify other types of interventions, such as quotas or affirmative action.
For example, many societies have historical ethnic or racial injustices.
In our terminology, we may say there were historical ethnic or racial biases.
Because of heritable economic, societal and cultural characteristics and circumstances, these historical biases may still impact current societal conditions.
Clearly, it is impossible to correct historical injustices by addressing the original biases because we cannot undo the past.
Introducing quotas or affirmative action might be one of the few ways such historical injustices can be addressed.

These considerations are related to our earlier discussion of discrimination (see Section~\ref{subsec:discrimination}).
Like the definition of discrimination used by the \citet{national_research_council_measuring_2004}, legal definitions of discrimination extend beyond our definition of bias.
In particular, legal definitions of discrimination also encompass what is called ``disparate impact''\footnote{\url{https://www.justice.gov/crt/fcs/T6Manual7}} in the US and ``indirect discrimination''\footnote{\url{https://eur-lex.europa.eu/EN/legal-content/summary/the-principle-of-equal-treatment-between-persons.html}} in the EU.
In the context of racial discrimination, this refers to a facially neutral, but insufficiently justified, rule or practice that results in different racial outcomes.
This is related to our definition of disparity.

Taking our definition of bias as a starting point, the question should not be whether there are different racial outcomes as a result of some rule, but whether that rule itself is justified.
For example, unjustified factors used in hiring should be disallowed, not because they result in racial differences per se, but because they are unjustified.
Suppose that only people with sufficient capital are considered for a certain job position.
On average, black people in the US have accumulated less capital than white people \citep{pager_sociology_2008}.
Hence, the requirement of having sufficient capital results in racial differences.
However, the requirement should be considered unjustified not because it results in racial differences, but because it is irrelevant to the job position.
Indeed, the unjustified requirement may have unfair results not only for black people, but also for other people who had fewer opportunities to accumulate capital.

Moreover, even if a rule does not result in any racial difference, it should be disallowed if it is considered unjustified.
In the above example, the requirement of having sufficient capital should be disallowed even if it does not produce any racial differences.
On the other hand, rules that result in racial differences may be allowed if the rules are justified.
For example, requiring educational qualifications in hiring procedures may be allowed even if it results in racial differences.
What is justified is of course context-dependent.
A requirement of having sufficient capital may be considered justified in the context of loans, while a requirement of educational qualifications may be considered unjustified in the context of voting rights.

Although discussions about discrimination law are far richer than what we cover here \citep{khaitan_theory_2015}, we believe that the ideas behind our concepts of bias and disparity can enrich these discussions.

\section{Case studies}
\label{sec:case_studies}

\subsection{Gender bias in science}
\label{subsec:gender_bias}

\noindent
Gender differences in science have been extensively studied in the literature, with differences observed in terms of publications, citations, funding and academic positions.
There are clear gender differences in citations \citep{Lariviere2013-bm}.
Some interpret these differences in citations as a gender bias, sometimes explicitly labelled as an implicit bias \citep{Dworkin2020-gp,Teich2021-yw}.
As a result, some suggest that this gender bias may be addressed by tools that check whether references are gender balanced, such as the Gender Balance Assessment Tool \citep{Sumner2018-lt}.
A possible explanation for gender differences in citations may be gender differences in seniority: there are often more men than women in more senior positions.
Some research corroborates this explanation and finds that gender differences in citations seem to result from gender differences in academic career trajectories and publication productivity \citep{Huang2020-pd}.
Another study attributes gender differences in citations to gender differences in journal prestige and collaboration patterns \citep{Andersen2019-vz}.
There are also observations of gender differences in self-citation rates \citep{King2017-to}, but this turns out to be mostly due to gender differences in publication productivity \citep{Mishra2018-oe}.

Several findings suggest that gender differences in publication productivity may explain other gender differences.
Gender differences in publication productivity have been termed a `productivity puzzle' in earlier literature \citep{Cole1984-qx}.
Some research suggests that articles authored by women are reviewed differently than those authored by men, with a shift from single-anonymous to double-anonymous peer review attenuating gender differences \citep{Budden2008-ep}.
However, other studies have found no such gender differences when comparing single-anonymous and double-anonymous peer reviews \citep{Blank1991-he,Tomkins2017-lb}.
A recent study suggests that gender differences in publishing do not emerge as a result of being reviewed differently \citep{Squazzoni2021-sv}, although that study's results might be affected by desk rejections \citep{Hagan2020-hx}.
Although family formation and related childcare responsibilities may provide an explanation, early studies find no supporting evidence \citep{Cole1987-bu}.
There may be relevant field differences: some fields may have more intense periods for career progression around the time of family formation, while other fields may demonstrate such more intense periods at other times \citep{Adamo2013-zt}.
In math-intensive fields, family formation is considered a key factor explaining gender differences \citep{Ceci2011-rv}.
Preliminary results from a large-scale survey suggest that women scientists indeed take on a caregiver role more frequently than men scientists, although the implications for productivity are not clear \citep{Derrick2021-ot}.

Women seem to transition into more senior positions less frequently than men, which may be partly explained by gender differences in publication productivity \citep{Lerchenmueller2018-kk}.
Although this is sometimes portrayed as a `leaky pipeline', there seems to be a particular point in this pipeline at which these gender differences are most pronounced, namely, the transition from postdoc to principal investigator \citep{Lerchenmueller2018-kk}.
After this transition, men and women seem to show similar career trajectories \citep{Kaminski2012-qn}.
There is evidence that men and women are evaluated differently when applying for academic positions, even with identical curricula vitae \citep{Steinpreis1999-fg}.
However, there is also evidence to the contrary \citep{Carlsson2020-re}.
This suggests that there is a gender disparity around the transition from postdoc to principal investigator, but whether this represents a gender bias in hiring or a gender bias at other steps in the causal pathway remains unclear.

Receiving funding is an important factor in making the transition from postdoc to principal investigator.
Some experimental evidence suggests that gender identities on funding applications do not produce gender differences in funding outcomes \citep{Forscher2019-eq}.
Other research suggests that gender differences in funding outcomes may depend on the criteria used to evaluate funding applications \citep{Witteman2019-nf}.
An analysis of Dutch data suggests gender differences in funding rates \citep{Van_der_Lee2015-dn}, but these differences may be confounded by the field of science \citep{Albers2015-cr}.
In a large review of the literature on gender differences in funding, \citet{Cruz-Castro2020-ox} observe that few studies in the area use an explicit causal framework, making it more difficult to draw rigorous conclusions.

The following subsection considers one study on the role of gender in mentorship to illustrate the importance of using an explicit causal framework.

\subsubsection*{Mentorship}

\noindent
As mentioned earlier, misinterpreting a gender difference as a gender bias or a gender disparity complicates matters.
Not all observed differences represent causal effects, with conditioning on a collider representing one particularly difficult problem.
A good example of this problem arises in a recent paper by \citet{AlShebli2020-mw} about the role of gender in mentorship.
The authors report that protégés with women mentors show a lower citation impact than protégés with men mentors.
This paper's publication engendered considerable debate, eventually resulting in its retraction.
Critics of the paper raised concerns about, for example, the data\footnote{\url{https://danieleweeks.github.io/Mentorship}} and the operationalisation of the concept of mentorship \citep{Lindquist2020-fd}, among other concerns.
In addition to these issues, we suggest that the analysis by \citet{AlShebli2020-mw} may suffer from conditioning on a collider.

\begin{figure}
  \includegraphics{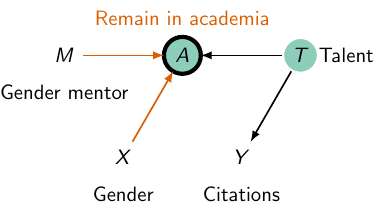}
  \caption{Simplified causal model of the role of gender in mentorship.}
  \label{fig:mentor}
\end{figure}

In Fig.~\ref{fig:mentor}, we present a simple causal model describing mechanisms relevant to interpreting the results of \citet{AlShebli2020-mw}.
According to our model, a person's research talent $T$ affects both the citations $Y$ they receive and their likelihood of remaining in academia $A$.
Independently of this, a person's gender $X$ and the gender of their mentor $M$ also affect their likelihood of remaining in academia.
More specifically, we assume that having a woman rather than a man mentor makes it more likely for a woman protégé to remain in academia \citep{Hofstra2022-wm}.

In our causal model, remaining in academia $A$ is $d$-connected to citations $Y$ because of the path $A \leftarrow \innode{T}{open} \rightarrow Y$, where talent $T$ acts as a confounder.
This is the only path (longer than a single link) that is $d$-connected.
All other paths are closed by node $A$, which acts as a collider for these paths.
Hence, citations $Y$ are independent of both the gender of the protégé $X$ and the gender of the mentor $M$.
Although it is debatable whether this is a realistic aspect of our model, our goal is not to construct a fully realistic model but to illustrate the potential problem of conditioning on a collider.

\citet{AlShebli2020-mw} make an explicit selection of the protégés included in their data collection: `we consider protégés who remain scientifically active after the completion of their mentorship period' (p.\ 2).
In our causal model, this amounts to conditioning on remaining in academia $A$ because this variable is used to select the protégés into the data set.
Conditioning on remaining in academia $A$ opens a number of paths that were previously closed, increasing the number of pairs of $d$-connected nodes.
For example, gender $X$ becomes associated with citations $Y$ because of the path $X \rightbias \innode{A}{open,conditioned} \leftarrow \innode{T}{open} \rightarrow Y$.
Moreover, the gender of the mentor $M$ becomes correlated with the citations $Y$ of the protégé because of the path $M \rightbias \innode{A}{open,conditioned} \leftarrow \innode{T}{open} \rightarrow Y$.
That is, there is a gender difference in citations for both the gender of the protégé and the gender of the mentor.
In our model, women protégés with men mentors are less likely to remain in academia, which means that those who do remain in academia can be expected to be more talented, on average, than their colleagues with women mentors.
Consequently, for protégés who remain in academia, having a woman mentor relates to being less talented, which leads to fewer citations.
Importantly, the association between citations and the gender of a protégé's mentor does \emph{not} reflect a causal effect.
Instead, it is the result of conditioning on a collider.
This example illustrates how conditioning on a collider easily leads to incorrect conclusions.
Depending on the extent to which our model captures the relevant causal mechanisms, the main result of \citet{AlShebli2020-mw} may be due to conditioning on a collider.

Our hypothetical model challenges the policy recommendations made by \citet{AlShebli2020-mw}.
The authors suggest that women protégés should be paired with men mentors because this positively affects their citation impact.
If our causal model holds true, this suggestion is incorrect.
In our model, pairing a woman protégé with a man mentor reduces the likelihood that the protégé remains in academia, meaning that protégés who do persevere in academia are likely to be more talented and obtain more citations.
The difference between men and women mentors in terms of the citations received by their protégés represents a gender difference and not a gender bias or gender disparity.
Without additional evidence or assumptions, the observed gender difference does not support the policy recommendations made by \citet{AlShebli2020-mw}.
In fact, given our conjectured model, it can be argued that one should do the opposite of what they suggest: to increase women participation in science, women protégés should be paired with men mentors.
This illustrates the importance of considering the appropriate causal mechanisms for making policy recommendations.

\subsection{Racial bias in police shootings}
\label{subsec:police_shooting}

\noindent
Police shootings in the United States are much more frequent than police shootings in Europe\footnote{\url{https://www.theguardian.com/us-news/2015/jun/09/the-counted-police-killings-us-vs-other-countries}}.
The overwhelming availability of guns coupled with a militarised police culture constitutes a deadly combination \citep{Hirschfield2015-ah}.
Additionally, recurrent concerns over racial biases in policing can be seen in the context of a long history of institutionalised racism in the United States \citep{Kendi2017-mb}.
In the past decade, multiple police killings of innocent Black people have led to large protests and sparked the Black Lives Matter movement.
Several newspapers have started collecting data about US police shootings and their victims, including \textit{The Guardian}\footnote{\url{https://www.theguardian.com/us-news/series/counted-us-police-killings}} and \textit{The Washington Post}\footnote{\url{https://www.washingtonpost.com/graphics/investigations/police-shootings-database/}}.
The academic literature analysing racial biases in police shootings has adopted these databases alongside other data collection efforts.

One highly publicised paper reported rather unexpected findings \citep{Fryer2019-ws}.
As expected, the paper observed racial bias against Black people in cases of non-lethal force by police in the United States.
However, the author found no evidence of racial bias against Black people in cases where the police used lethal force.
In fact, he reported finding that Black people were less likely than White people to be shot in a police encounter.
This paper caused quite a stir and received widespread attention from academics and the popular press when it was published as an NBER working paper.
Shortly afterwards, several researchers began critically examining the results of \citet{Fryer2019-ws}, with the findings called into question by several blog posts and academic articles \citep{Ross2018-bp,Knox2020-at}.

\begin{figure*}
  \includegraphics{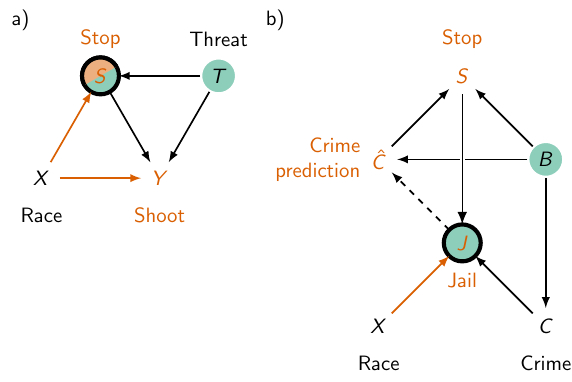}
  \caption{Hypothetical causal models of (a) racial bias in police shootings and (b) racial bias in jailing people.}
  \label{fig:shooting_bias}
\end{figure*}

A central point of critique is the causal aspect of the study, which we summarise in a simple causal model in Fig.~\ref{fig:shooting_bias}a.
In particular, \citet{Fryer2019-ws} analysed the probability that a suspect was shot, which implies that the probability was conditional on the suspect being stopped.
Presumably, police are more likely to stop people who pose some threat, that is, $T \rightarrow S$.
If there is racial bias against Black people in stops by police, $X \rightbias S$, then Black people who are stopped tend to be less threatening than non-Black people, induced by conditioning on the collider $S$.
Indeed, \citet{Gelman2007-vi} reported that Black people are more likely to be stopped relative to the general population, even after controlling for crime rate differences.
If we assume that the threat level influences the probability that police will shoot, or $T \rightarrow Y$, then, conditional on being stopped, Black people are less likely to be shot because Black people who are stopped are less likely to be a real threat.
Although there might still be an actual bias against Black people that counteracts this effect, $X \rightbias Y$, it might not outweigh the effect of conditioning on the collider of being stopped.

The difficulty is that we cannot identify the direct causal effect of race on being shot unless we also condition on being a threat.
This means that \citet{Fryer2019-ws} did not correctly identify racial bias in police shootings, and a racial bias in police shootings cannot be established based on those results.
That is, the results of \citet{Fryer2019-ws} do not clarify whether the arrow $X \rightbias Y$ is present in the causal model in Fig.~\ref{fig:shooting_bias}a.
Regardless of the potential racial bias in police shootings $X \rightbias Y$, the racial bias in stopping $X \rightbias S$ would imply a racial disparity in police shootings.

This resembles the argument advanced by \citet{Knox2020-at}: if police are biased in stopping citizens, any study that uses records of stopped citizens will not be able to correctly infer the causal effect of being shot.
According to \citet{Knox2020-at}, under some conditions, the total causal effect of race on being shot can be estimated.
However, the direct causal effect is impossible to estimate without making very strong assumptions.
Our simple causal model (Fig.~\ref{fig:shooting_bias}a) also reveals that if we condition on $S$, we condition on a collider, thereby opening a path of confounding influence of $T$. 
However, if we do not condition on $S$, we leave open a mediating path.
Either way, we cannot identify the direct causal effect of race on being shot unless we can credibly measure and correct for the threat level $T$.
We previously encountered this structural causal model in Fig.~\ref{fig:example_collider} in our discussion of problems with identifying biases.
Indeed, \citet{Knox2020-at} call for more attention to the issue of causality, emphasising that without clearly understanding the causal structure, it is difficult to formulate effective policy.

\citet{Fryer2018-xb} aims to reconcile various outcomes, considering not only his earlier work \citep{Fryer2019-ws}%
    \footnote{\citet{Fryer2019-ws} appeared as an NBER working paper in 2016 and was published as a journal article in 2019, while \citet{Fryer2018-xb} appeared as a conference paper in 2018 (before \citet{Fryer2019-ws} was published as a journal article).
    Although it might seem that \citet{Fryer2018-xb} cited future work, this is not the case.}%
and work by \citet{Ross2015-uw} but also various data sources from newspaper outlets including \textit{The Washington Post}, \textit{The Guardian} and VICE.
\citet{Fryer2018-xb} acknowledges that the data largely paint a similar picture, namely of large racial disparities in police shootings.
\citet{Fryer2018-xb} argues that his earlier research \citep{Fryer2019-ws} is unique because it controls for factors that others do not control for.
However, as explained, the controls used by \citet{Fryer2019-ws} do not enable the identification of a bias in police shootings.
\citet{Fryer2018-xb} seems to suggest that if there is no bias in police shootings, there is no problem.
However, from our perspective, even if there is no bias in police shootings, there may still be a disparity in police shootings caused by a bias in police stoppings, and such a disparity would still be problematic.
In this situation, it would not be possible to correct the disparity by addressing a bias in police shootings, but the disparity could be corrected by addressing the bias in police stoppings.

\citet{Ross2018-bp} also aim to reconcile the results of \citet{Fryer2019-ws} and \citet{Ross2015-uw}.
Using a formal Bayesian model, they show that a racial disparity in population-level police shootings may be observed even if there is no racial bias in the shootings themselves.
The disparity may be due to a racial bias in stopping.
If police officers are more likely to stop Black citizens, these citizens will be less likely to pose a threat and will, therefore, be less likely to be shot.
Again, this is similar to the previously discussed problem of conditioning on a collider.

For \citet{Ross2018-bp}, the total causal effect is the relevant public health perspective.
From the viewpoint of identifying the problem, we agree.
However, from the viewpoint of addressing the problem, this may be insufficient.
If racial disparities in police shootings stem from a racial bias in police shootings, policies aimed at addressing this bias---such as use-of-force training, stricter enforcement of police shooting regulations and redrafting such regulations---may be helpful.
In contrast, if racial disparities in police shootings stem from a racial bias in encounter rates, different policies are needed.
For example, if there is a racial bias in patrolling intensity, resulting in more intense policing in neighbourhoods with more Black people, policies aimed at addressing racial bias in police shootings are unlikely to be effective, and revising patrolling intensity may represent a wiser strategy.
This aligns with further arguments made by \citet{Ross2018-bp}.

Finally, \citet{Cesario2021-az} discusses the challenges associated with using experimental studies of racial bias in police shootings to inform observational studies.
He argues that it is difficult to translate the results of experimental studies directly to real-world settings.
This is more generally known as the problem of transportability \citet{Pearl2014-ss}, as also suggested by \citet{Rohrer2021-ht}.

\subsubsection*{Benchmarks and statistical discrimination}

\noindent
\citet{Fryer2019-ws} raises the issue of a `risk-set', or what \citet{Cesario2019-qr} call a `benchmark'.
This discussion has two sides, both of which are relevant to the discussion of bias and disparity: causality and the question of justification.
Although there is an over-representation of Black people in police shootings relative to the overall population, according to \citet{Cesario2019-qr}, there is no such over-representation relative to the population engaged in criminal activity.
Yet using criminal activity as the benchmark does not necessarily identify a direct causal effect or a bias, although it may identify a disparity.
\citet{Ross2021-dq} use a formal model to show that separating armed and unarmed people who were shot may offer a benchmarking approach that allows identifying a causal effect, that is, a disparity.
This requires assuming that armed people are from a criminal population, unarmed people are from a non-criminal population, and arrest rates reflect rates of criminal activity.
Although the assumption that armed people are from a criminal population may be warranted, the assumption that unarmed people are from a non-criminal part of the population is less convincing, as \citet{Cesario2020-ew} also argued.
Contrary to the argument by \citet{Cesario2020-ew}, this invalidates both the `benchmark correction' used for unarmed people and the `benchmark correction' used for armed people.
In short, formal modelling may help uncover whether a `benchmark' does or does not lead to correct causal estimates.
However, such causal estimates most likely do not reflect a direct causal effect and do not identify a bias.

Some may argue that a direct causal effect of race on stopping, $X \rightarrow S$, is justified because of differences in crime rates across racial groups, $X \rightarrow C$.
According to this argument, patrolling certain crime-intensive areas more often or being called to a scene more frequently are the result of crime rate differences across racial groups.
A person's race may be predictive of the probability that a situation will involve criminal activity and, consequently, that an individual should be stopped.
Hence, different stopping rates for different racial groups may be justified because of different rates of involvement in criminal activities.
\citet{Fryer2019-ws} refers to this as statistical discrimination, which we discussed in Section~\ref{subsec:discrimination}.

To fully appreciate the complexity of this situation, a proper understanding of causality is essential.
This also connects to the discussion of fairness in AI in Section~\ref{sec:fairness}.
Suppose that there is no direct causal effect of race $X$ on crime $C$, as the causal model in Fig.~\ref{fig:shooting_bias}b illustrates.
However, suppose also that there is a racial bias in being jailed for a crime, $X \rightbias J$.
That is, whether someone is jailed $J$ depends not only on the crime $C$ they committed but also on their race $X$.
Additionally, suppose that there are certain behavioural features $B$ that influence whether a person commits a crime $C$.
Because these behavioural features $B$ are independent of race $X$, there is no causal effect of race $X$ on crime $C$.
Now, suppose we want to predict crime $\hat{C}$.
If we only predict crime $\hat{C}$ based on $B$, $\hat{C}$ is a fair predictor and shows no racial bias or disparity.
However, if we also consider having been in jail $J$ in the prediction, $\hat{C}$ suffers from a racial disparity.

Studies that analyse the convicted population, as in the case of the `benchmark' approaches \citep{Cesario2019-qr}, condition implicitly on $J$.
Because $J$ is a collider, $X \rightbias \innode{J}{open,conditioned} \leftarrow C$, race $X$ becomes predictive of crime $C$, even though there is no causal effect of race on crime.
When the prediction $\hat{C}$ is actually used in practice, it leads to a feedback loop. 
That is, $\hat{C}$ will affect the stopping rate $S$, which in turn affects whether a person is jailed $J$.
When analysing the convicted population a few years after introducing the prediction $\hat{C}$, racial differences in crime will again become apparent, seemingly confirming the correctness of the prediction $\hat{C}$.
However, this prediction is correct precisely because it exhibits a racial disparity that influences whether someone is stopped $S$ and jailed $J$.
It does not reflect a causal effect of race $X$ on crime $C$.
Hence, the racial bias in being jailed $J$ leads to downstream racial disparities, disparities that then have a self-reinforcing effect.
This is a prime example of what \citet{ONeil2016-sa} calls a pernicious feedback loop.
Even when using AI `just for prediction', whenever we act upon a prediction, we enter causal territory.
Such pernicious feedback loops are not limited to AI and may operate equally well in social processes such as the one discussed here.

\section{Discussion}
\label{sec:discussion}

\noindent
We propose definitions of bias and disparity using the framework of structural causal models \citep{Pearl2009-vd}.
We define a bias as a direct causal effect that is unjustified.
A bias involves both a variable that acts as the cause and a variable that acts as the outcome.
For example, in the case of a gender bias in hiring, an individual's gender is the cause and whether that individual is hired is the outcome.
Whether a certain causal effect is considered justified is an ethical question that cannot be determined empirically on the basis of data.
We consider disparity a broader concept than bias.
We define a disparity as a direct or indirect causal effect that includes a bias.
There is a disparity of $X$ in $Y$ if at least one link on a causal pathway from $X$ to $Y$ represents a bias.
If there is a disparity of $X$ in $Y$, the outcome $Y$ is considered unfair with respect to $X$.

It is important to understand whether an observed difference represents a bias, a disparity or neither.
If a difference does not represent a bias or disparity, there is probably no need for a policy intervention.
Differences that represent a bias or disparity offer grounds for intervention.
When intervening to correct a disparity, it is important to know where in the causal pathway the bias is located.
If possible, the intervention should aim at correcting the bias.
Although interventions elsewhere in the causal pathway (e.g. affirmative action) may occasionally be deemed necessary, they do not solve the fundamental problem of the bias.
Without a proper causal understanding, we risk making incorrect policy recommendations, which may even yield an outcome that is completely contrary to the intended outcome.

The notion of fairness has been discussed extensively in the AI literature \citep{Oneto2020-fh}.
Several popular fairness criteria in AI are incompatible with the notion of fairness underlying our definitions of bias and disparity.
These AI-based fairness criteria classify predictions as fair that are considered unfair according to our definitions and vice versa.
Unlike the AI-based fairness criteria analysed by this paper, our definitions acknowledge that fairness requires an ethical judgement and cannot be determined using data alone.
Counterfactual fairness \citep{Chiappa2019-yz} is a promising approach to fairness in AI that closely relates to our definitions of bias and disparity.
Whether AI techniques can be applied without reproducing existing biases in data depends on our causal understanding of the data at hand.
By properly understanding causality, AI might reduce biases that currently prevail in the real world.
However, simplistic fairness definitions that ignore causality are problematic and likely to perpetuate biases.
Moreover, using AI `just for prediction' is no escape: whenever we act upon a prediction, we enter causal territory.

Notably, explicit definitions of the concepts of bias and disparity have rarely been provided in the literature.
We believe that our proposed definitions reflect the intuitive understanding that many researchers have of these concepts.
We hope that our definitions will help researchers to reason in more precise ways about biases and disparities, contributing to more consistency in the use of these concepts in future research.

\section*{Competing interests}
\noindent The authors declare no competing interests.

\section*{Funding information}
\noindent The authors acknowledge no funding for this paper.

\bibliographystyle{aipauth4-2-doi}
\nocite{apsrev41Control}
\bibliography{control,bibliography}

\end{document}